\documentclass[%
 reprint,
 amsmath,amssymb,
 longbibliography,
 aps,
 pre,
]{revtex4-1}

\usepackage[normalem]{ulem}
\usepackage{graphicx}
\usepackage{bm}
\usepackage{xcolor}
\usepackage{hyperref}
\usepackage{amsmath}
\usepackage[utf8]{inputenc}
\usepackage{ulem}
\usepackage{mathtools}
\usepackage{xr}
\usepackage{caption}
\usepackage{subcaption}
\usepackage{svg}
\usepackage{soul}

\begin{document}

\title{Numerically Discovered Inherent States are Always Protocol Dependent in Jammed Packings}
\author{Eddie Bautista}
\email{ebautis2@uoregon.edu}

\author{Eric I. Corwin}
\email{ecorwin@uoregon.edu}
\affiliation{Department of Physics and Materials Science Institute, University of Oregon, Eugene, Oregon 97403, USA. }
\date{\today}

\begin{abstract}
The energy landscape for soft sphere packings exists in a high-dimensional space and plays host to an astronomical number of local minima in a hierarchical and ultrametric arrangement. Each point in the landscape is a configuration that can be mapped to its inherent state, defined as the local minima that the configuration will flow to under perfectly overdamped continuous dynamics. Typically, discrete in time dynamics are used to computationally find local minima, but it is not known whether these algorithms are capable of reliably finding inherent states. Here, we use steepest descent dynamics to find the distribution of the largest time step, $\delta_\textrm{best}$, which finds the inherent state. We find that $\delta_\textrm{best}$ is Weibull distributed. Additionally, for systems of $N$ particles, $\delta_\textrm{best}$ falls rapidly with increasing $N$, with a functional form somewhere in between an exponential and inverse power-law, and is weakly dependent on $d$ and $\varphi$.  We argue that this rapid fall is due to saddle points in the energy landscape. Our results suggest that it is impossible, in practice, to reliably find inherent states for systems of about 64 particles or more. 
\end{abstract}

\maketitle


\textit{Introduction}---Granular packings of soft spheres exist in a low-density unjammed phase and a high-density jammed Gardner phase \cite{berthier_gardner_2019,castellani_spin-glass_2005}. Separating them is the jamming transition, a random first-order, athermal phase transition~\cite{liu_jamming_1998, berthier_equilibrium_2016, charbonneau_glass_2017, rintoul_metastability_1996, castellani_spin-glass_2005, charbonneau_finite-size_2021, arceri_jamming_2022, bertrand_protocol_2016}. The energy landscape in the low-density phase is simple and smooth, while that of the high-density phase consists of a fractally rough set of hierarchical sub-basins within sub-basins \textit{ad infinitum} playing host to an astronomical number of local minima~\cite{artiaco_exploratory_2020,charbonneau_exact_2014, charbonneau_glass_2017, dennis_jamming_2020, charbonneau_fractal_2014-1, thirumalaiswamy_exploring_2022, rainone_following_2015,urbani_shear_2017}.

Each point in the high-density landscape can be unambiguously mapped to its ``inherent state'', which is defined as the local minima that the configuration will flow to under perfectly overdamped continuous dynamics~\cite{sciortino_potential_2005, martiniani_when_2023,speedy_random_1998, martiniani_structural_2016, gao_enumeration_2007}. These dynamics are often approximated with various discrete in time dynamics in order to computationally approximate the inherent state. This mapping segments the landscape into individual basins of attraction~\cite{martiniani_when_2023, ashwin_calculations_2012}. This segmentation of space allows one to define the Shannon entropy for the landscape, which is expected to control thermodynamic and statistical mechanical properties of granular packings~\cite{martiniani_when_2023, martiniani_numerical_2017}. However, it is not known whether such discrete time dynamics can reliably find the true inherent state.

Here, we sample the energy landscape and flow to local minima using steepest descent dynamics with systematically varying time steps. We find that for every starting configuration, there exists some time step above which the system falls into a minima other than the inherent state. We find that the value of this time step needed to find the inherent state decays rapidly with particle number and is only weakly dependent on packing fraction, $\varphi$, and spatial dimension, $d$. We argue that the strong correlation between time step and particle number is due to saddle points in the fractally rough energy landscape.  We find that these constraints place limits on the computability of inherent states, under any circumstances, to systems of about $N=64$ particles or fewer.

Many discrete numerical algorithms exist to find local minima in the landscape. Some examples include CVODE~\cite{hindmarsh_sundials_2005, gardner_enabling_2022}, FIRE~\cite{bitzek_structural_2006, parisi_theory_2020}, L-BFGS~\cite{liu_limited_1989}, and Steepest Descent~\cite{cauchy_methode_nodate}. Some momentum-based algorithms, such as FIRE and L-BFGS, have already been found to incorrectly identify inherent states due to the introduction of an effective ``noise'' in the energy landscape~\cite{suryadevara_basins_2025, charbonneau_jamming_2023}. This noise causes the packings to jump over small barriers in the landscape into neighboring minima. In this work, we show that even first-order iterative algorithms such as Steepest Descent are only able to find the inherent state when using extremely small step sizes. We use the steepest descent minimizer because it is the simplest and best understood minimizer, thus ensuring any behavior we find is due to the energy landscape and not the complexities of the minimizer. 

\begin{figure}
\centering
\includegraphics[width=\columnwidth]{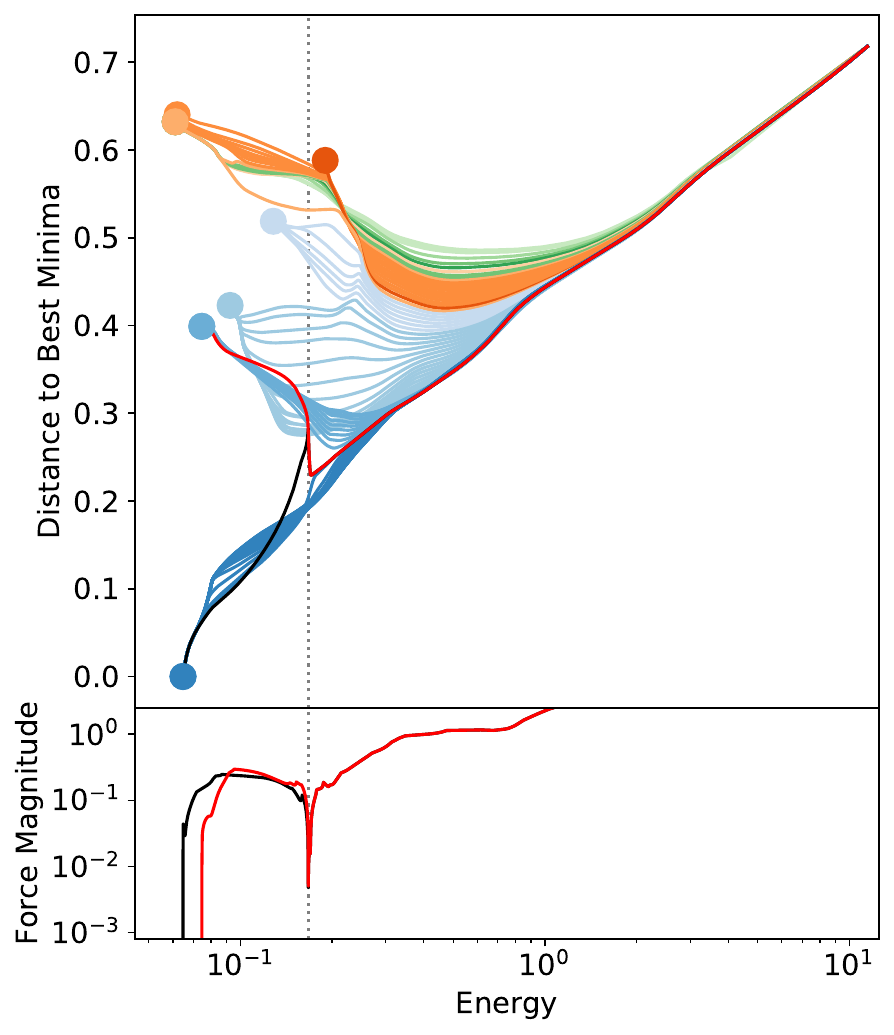}
\caption{(Top) Steepest descent minimization paths using varying time steps starting from the same initial state. We minimize the initial packing using time steps $\delta$ between $10^{-6}$ and $10^{-2}$. Each line represents the SD path of a different $\delta$, colored according to the unique minima to which the state will flow. The final states are indicated by circle markers. We see the curves naturally segment themselves based on the time step. The black and red curves show the path for $\delta$'s right above and right below $\delta_\textrm{crit}$ for the saddle point separating the best minima and its first neighbor. We see the paths stick closely together until the saddle point energy, at which they separate sharply. (Bottom) Force plot for the black and red curves. At the point of separation, indicated by the black dotted line, we see a drastic dip in the force. This tells us that it is a saddle point in the energy landscape.}
\label{saddle}
\end{figure}
%


\textit{Model}---We simulate $N$ harmonic soft spheres in a $d$-dimensional box of side length 1 with periodic boundary conditions. The total energy of a system composed of spheres with locations $\{\vec x_i\}$ and radii $\{r_i\}$ is defined as 
\begin{equation}
U(\{\vec x_i\}) =\frac{1}{2} \sum_{ij} \left ( 1-\frac{ \| \vec{x}_i-\vec{x}_j\|}{r_i+r_j} \right ) ^2 \Theta \left( 1-\frac{ \| \vec{x}_i-\vec{x}_j\|}{r_i+r_j} \right) ,
\end{equation}
where $\Theta$ is the Heaviside function necessary for a contact potential.

The dynamics to reach an inherent state are defined as 
\begin{equation}
    \frac{\partial \{\vec x_i\}}{\partial t} = -\nabla U(\{\vec x_i\}).
\end{equation}
We discretize these dynamics to form a steepest descent (SD) minimizer with a fixed time step, $\delta$, as
\begin{equation} \label{eqn:SD}
\{ \vec{x}_{i,t+1} \} = \{ \vec{x}_{i,t} \} - \delta \nabla U(\{x_{i,t}\}),
\end{equation}
where $\{\vec x_i,t\}$ is the position of all particles at time step $t$. Throughout the minimization, $\delta$ is kept constant.


\textit{Methods}---An energy landscape is uniquely defined by the packing spatial dimension, $d$, particle number, $N$, and the set of particle radii, $\{r_i\}$.  To prevent crystallization in $d$=2, we use radii drawn from a log-normal distribution with a polydispersity (ratio of standard deviation to mean) equal to 0.25. In higher dimensions, we use monodisperse radii, as there is no spontaneous crystallization. For a given landscape, we sample 1000 random initial states uniformly across the space.

For each initial state, we minimize the packing energy using the steepest descent dynamics in Eq.~\eqref{eqn:SD} as implemented by the \textsc{pyCudaPacking} software~\cite{morse_geometric_2014, charbonneau_universal_2016, morse_echoes_2017}, a GPU-based sphere packing minimizer. We use time steps ranging from $\delta = 10^{-6}$ to $10^{-2}$ in natural units. We chose $10^{-6}$ as our lower limit because of time and resource constraints. Steps larger than $10^{-2}$ tend to overshoot minima or diverge at high rates, therefore making them unsuitable for exploration.

For each choice of $\delta$, a given initial configuration can, in principle, minimize to a different final state in the energy landscape. However, many values of $\delta$ will minimize the initial state to the same minimum, as shown in Fig.~\ref{saddle}. To determine whether two minima are the same, we find that it is sufficient to compare their energies~\cite{hagh_order_2024}; if the energies differ by less than $10^{-10}$ we can be confident that the systems are in the same minima. In $d=2$, we verify this by looking at the contact vector distance to identify minima, as was done in reference~\cite{dennis_jamming_2020}. This verification only works in $d=2$ because the radii are polydispersed; thus, the particles are distinguishable.

Ideally, we would compare the minima found with a given value of $\delta$ to the true inherent state. However, lacking any scheme to produce such an inherent state, we instead define the ``best minima'' as the minima found using the smallest practicable time step, $10^{-6}$. 

For a given starting point, we determine $\delta_\textrm{best}$, the largest time step such that all smaller time steps result in the best minima. The value $\delta_\textrm{best}$ gives us insight into how small a time step is needed to be reasonably sure we have found the inherent state. To see how $\delta_\textrm{best}$ changes, we sample packings in $d=2$, 3, and 4 spatial dimensions with different particle numbers N, and packing fractions $\varphi$. 


\textit{Results}---In order to understand how the discrete time dynamics influence the trajectory of a system, we look at the steepest descent path. For a given starting configuration, we first find the best minima using $\delta = 10^{-6}$. Then, for a range of values of $\delta$ logarithmically spaced up to $10^{-2}$, we record, at every minimization time step, the energy, force magnitude, and distance in position space to the best minima.

Fig.~\ref{saddle} (top) shows these discrete time steepest descent paths for a $d=2$ packing of $N=16$ particles at $\varphi=1.07\varphi_J$, where $\varphi_J$ is the jamming packing fraction. Each line represents a different time step, and the color represents into which unique minima the packing minimizes. We see that for time steps lower than $\sim5\times10^{-3}$, the lines are well segmented by color, meaning that there is a continuous range of $\delta$ for which the SD dynamics all minimize to the same minima. The value of $\delta$ which separates each adjacent range of time steps is a critical timestep, $\delta_\textrm{crit}$, whose dynamics bring the system not to a minima, but to a saddle point which separates those two minima.

\begin{figure}[t!]
\centering
\includegraphics[width=\columnwidth]{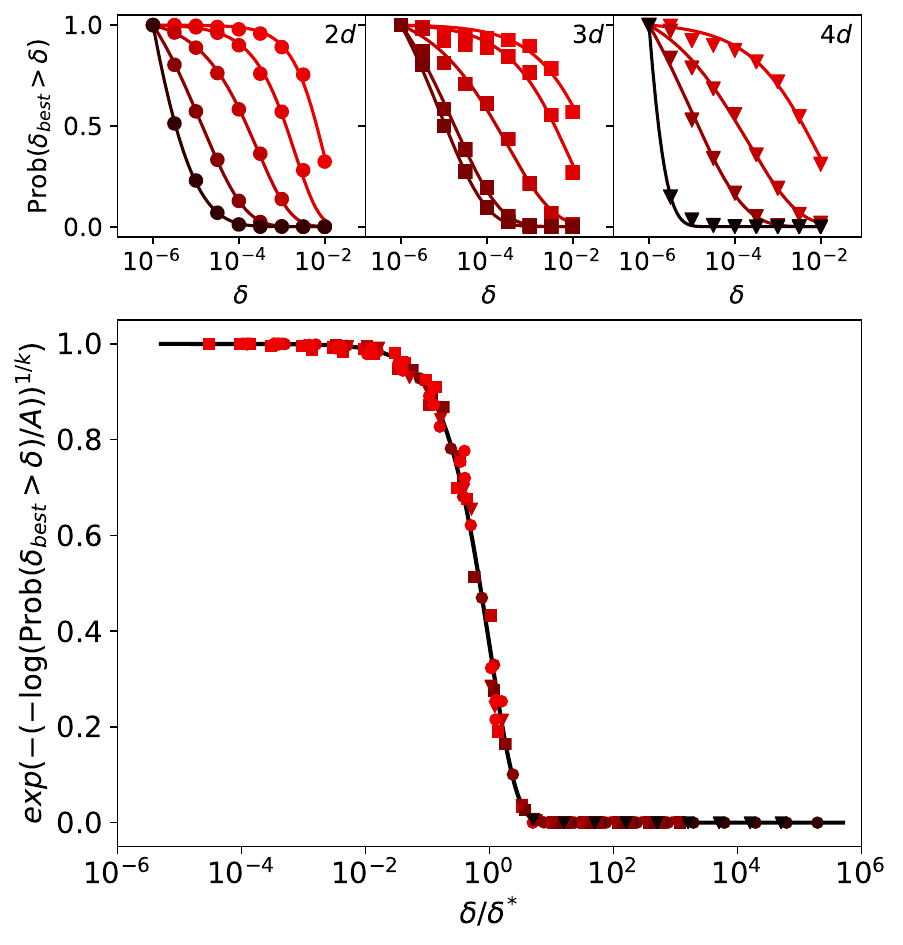}
\caption{
(Top) Complementary cumulative distribution function of $\delta_\textrm{best}$ for packings in $d=2$ (left), $d=3$ (middle), and $d=4$ (right) spatial dimensions with changing $N$ and $\varphi$ held constant at $1.07 \varphi_J$. The y-axis shows the probability that $\delta_\textrm{best}$ is above a given time step $\delta$. The colors represent different $N$ with darker colors conveying larger particle numbers. The lines show the fits to the complementary cumulative distribution function of the Weibull distribution, Eq.~\eqref{eqn:ccdfWeibull}. (Bottom) Scaled complementary cumulative distribution function for all data in the top figure. The x-axis is scaled by  $\delta^*$, and the y-axis has the A and k parameters scaled out. The circles represent the $d=2$ data, the squares $d=3$, and the triangles $d=4$. Also included are $d=2$, $N=16$ packings with $\varphi = 1.01 \varphi_J$ and $1.13 \varphi_J$. }
\label{CCDF}
\end{figure}

For energy landscapes in high-dimensional spaces, low-index saddle points are attractive in all but a small number of directions. Thus, the trajectory of a perfectly overdamped continuous dynamics will approach closely to each saddle point that it encounters on the way down to the inherent state. Further, in high dimensions, a random initial state is likely to be near a border between basins of attraction in the energy landscape and so the perfectly overdamped trajectory of such a state will follow the border towards a saddle \cite{kurchan_phase_1996}. The time step in discrete time SD dynamics introduces an effective momentum, which allows the system to deviate from the perfectly overdamped path. With enough momentum and close to a saddle point, the system can be carried over the separatrix between neighboring minima. Previous work has shown the importance of saddles in the energy landscape of glassy materials \cite{grigera_geometric_2002,cavagna_numerical_2004, cavagna_stationary_1998, cavagna_fragile_2001, cavagna_role_2001}.

We can verify these are indeed saddle points by first adjusting $\delta$ so that the minimization path comes arbitrarily close to the separatrix between wells. The black and red lines in Fig.~\ref{saddle} (top) are the paths for time steps bracketing $\delta_\textrm{crit}$ for a representative saddle point.  The red curve has $\delta = \delta_\textrm{crit} + 10^{-20}$ and the black curve has $\delta = \delta_\textrm{crit} - 10^{-20}$. We see that the paths are identical until they hit the saddle point, whose energy is indicated by the dashed vertical line, at which point they abruptly begin to separate.  Fig.~\ref{saddle} (bottom) shows the force vs energy plot of the black and red curves. At the saddle point, there is a sudden dip in the force towards zero, consistent with a critical point. Examining the Hessian reveals that it is an index one saddle point, meaning that there is a single repulsive eigenvector for the associated dynamical, or Hessian, matrix, and all other eigenvectors are attracted to the saddle point.

For a given starting configuration, we can thus see that
\begin{equation}
    \delta_\textrm{best} = \textrm{Min}[\{\delta_{\textrm{crit},i }\}],
\end{equation}
where $\{\delta_{\textrm{crit},i }\}$ is the set of saddle points encountered by the minimization path.
One should expect the resulting $\delta_\textrm{best}$ to be distributed according to a Weibull distribution~\cite{weibull_statistical_1951, majumdar_extreme_2020} with centering and width determined by the functional form of the lower tail of the distribution of $\delta_\textrm{crit}$. In fact, under certain circumstances, one might find a Gumbel distribution instead, but this proves a poor fit to our measured data and so is not explored here. When measured with respect to the true inherent state, we should then find $\delta_\textrm{best}$ distributed according to
\begin{equation} \label{eqn:weibull}
f(\delta;\delta^*,k)=\frac{k}{\delta^*}(\frac{\delta}{\delta^*})^{k-1} e^{-(\delta/\delta^*)^k}
\end{equation}
where k is the shape parameter and $\delta^*$ is the scale parameter. 

This picture is complicated by the fact that the measured value of $\delta_\textrm{best}$ is \textit{not} the value below which the system will always minimize to the true inherent state, but rather the value below which the system will always minimize to the same minimum as when $\delta=10^{-6}$.  This can be incorporated into a prediction for the complementary cumulative distribution of the measured $\delta_\textrm{best}$ as a truncated Weibull,
\begin{equation} \label{eqn:ccdfWeibull}
\textrm{Prob}(\delta_\textrm{best} > \delta) = Ae^{-(\delta/\delta^*)^k},
\end{equation}
where k and $\delta^*$ are the same shape and scale parameter as in Eq.~\eqref{eqn:weibull} and $A = \exp((10^{-6}/\delta^*)^k)$ is a normalization factor to enforce $\textrm{Prob}(\delta_\textrm{best} > 10^{-6}) = 1$.

Fig.~\ref{CCDF} (top) Shows the complementary cumulative distribution function of $\delta_\textrm{best}$ for $d=2$ (left), $d=3$ (middle), and $d=4$ (right) packings with  $\varphi = 1.07 \varphi_J$. The y-axis shows the probability of finding the best minima using time step $\delta$. Each color represents a different particle number, and each line shows the fits to the complementary cumulative distribution function. In $d=2$, $N = 16$, 32, 64, 128, 216. In $d=3$, $N=19$, 31, 64, 128, 144. Finally, in $d=4$, $N=32$, 64, 108, 256. For a given value of $N$ we find that this probability falls monotonically with increasing values of $\delta$. 

Additionally, we look at the distribution of $\delta_\textrm{best}$ for the $d=2$, $N=16$ packings at $\varphi = 1.01 \varphi_J$ and $1.13 \varphi_J$.  As shown in Fig.~\ref{CCDF} (bottom), every $d$, $N$, and $\varphi$ distribution can be collapsed onto a master curve by scaling the x-axis by $\delta^*$ and scaling out the A and k parameters from the y-axis. All data collapses well to this master curve, affirming that $\delta_\textrm{best}$ is indeed Weibull distributed.


%
\begin{figure}
\centering
\includegraphics[width=\columnwidth]{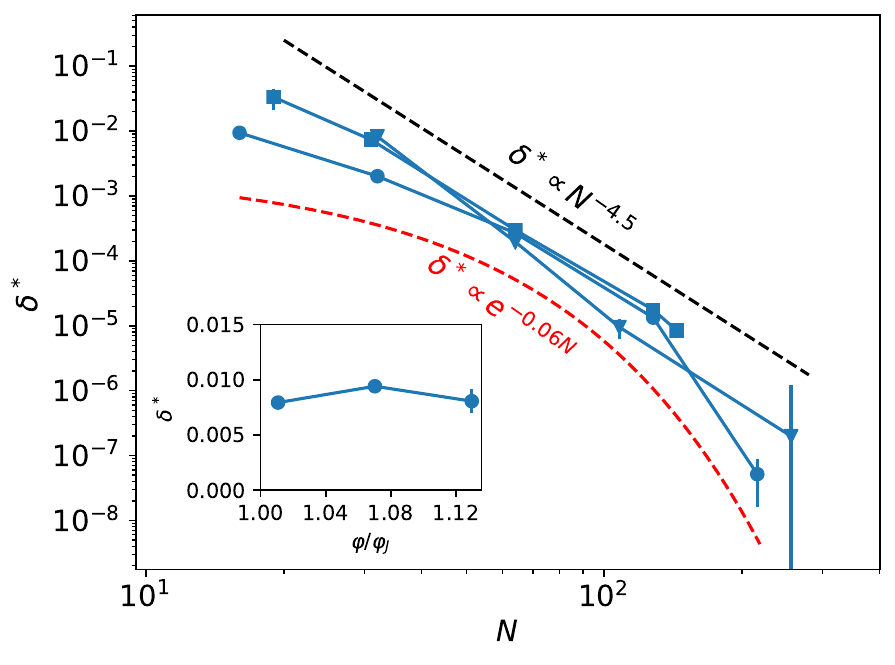}
\caption{Log-log plot of $\delta^*$ as a function of particle number for $d=2$, 3, and 4 spatial dimensions at packing fractions of $1.07\varphi_J$. The circles represent the $d=2$ data, the squares $d=3$, and the triangles $d=4$. The black dashed line shows a power law decay, while the red dashed line shows an exponential decay. Over this range, we see $\delta^*$ fall with a functional form in between an exponential and an inverse power law. In the inset, we show $\delta^*$ as a function of packing fraction for a $d=2$ packing with $N=16$. Here we see $\delta^*$ is nearly independent of $\varphi$}
\label{PowerLaw}
\end{figure}

Fig.~\ref{PowerLaw} shows $\delta^*$, the characteristic time-step of the distribution, as a function of particle number on a log-log plot for packings in $d=2$, 3, and 4 spatial dimensions. All packings are at a packing fraction of $1.07\varphi_J$. We see that over this, admittedly limited range in particle number, $\delta^*$ falls with a functional form somewhere in between an exponential and an inverse power law. Thus, there is a sharp drop-off in $\delta_\textrm{best}$ with increasing $N$, meaning we have to go to increasingly smaller time steps as we gradually increase the particle number. As a result, it is practically impossible to reliably find inherent states for systems of more than about 64 particles. 

Fig.~\ref{PowerLaw} inset, shows $\delta^*$ as a function of $\varphi$ for a $d=2$ packing with $N=16$. Here we see that $\delta^*$ is nearly independent of $\varphi$.



\textit{Conclusion}---These results demonstrate that it is virtually impossible to find inherent states for granular packings of large numbers of particles (in our case, 64 particles or more). This is because the steepest descent paths are attracted to saddle points in the energy landscape, each of which has an associated critical time step,  $\delta_\textrm{crit}$, above which the system will be shunted into the wrong minimum. In order to find the true inherent state, one must choose a time step, $\delta$, that is lower than the smallest $\delta_\textrm{crit}$. The set of time steps leading to inherent states are Weibull distributed and their centering falls rapidly with $N$, more quickly than a power-law but slower than an exponential. This centering is seemingly independent of $d$ and $\varphi$.

Adaptive minimization algorithms that change the time step as they minimize can speed up the beginning of the minimization process, thus allowing one to employ smaller time steps $\delta$. However, we still must go to extremely small time steps to find the inherent state. Most of the computational effort will be spent close to these saddle points and using very small time steps; therefore, the adaptive algorithms will likely have only a marginal effect on the computational resources needed to accurately find the inherent state for large systems.  Thus, the framework of the inherent state tessellation of space, while conceptually attractive, seems to be impractical. 


\textit{Acknowledgments}---We thank Cam Dennis, Peter Morse, Jorge Kurchan, Giorgio Parisi, Sean Ridout, and Ivan Corwin for useful discussions. This work was supported by a grant from the Simons Foundation No. 454939. The simulations were performed on the University of Oregon high-performance computing cluster, Talapas.

\bibliographystyle{unsrt}
\bibliography{references}

\begin{thebibliography}{10}

\bibitem{berthier_gardner_2019}
Ludovic Berthier, Giulio Biroli, Patrick Charbonneau, Eric~I. Corwin, Silvio Franz, and Francesco Zamponi.
\newblock Gardner physics in amorphous solids and beyond.
\newblock {\em The Journal of Chemical Physics}, 151(1):010901, July 2019.

\bibitem{castellani_spin-glass_2005}
Tommaso Castellani and Andrea Cavagna.
\newblock Spin-{Glass} {Theory} for {Pedestrians}.
\newblock {\em Journal of Statistical Mechanics: Theory and Experiment}, 2005(05):P05012, May 2005.
\newblock arXiv:cond-mat/0505032.

\bibitem{liu_jamming_1998}
Andrea~J. Liu and Sidney~R. Nagel.
\newblock Jamming is not just cool any more.
\newblock {\em Nature}, 396(6706):21--22, November 1998.
\newblock Publisher: Nature Publishing Group.

\bibitem{berthier_equilibrium_2016}
Ludovic Berthier, Daniele Coslovich, Andrea Ninarello, and Misaki Ozawa.
\newblock Equilibrium {Sampling} of {Hard} {Spheres} up to the {Jamming} {Density} and {Beyond}.
\newblock {\em Physical Review Letters}, 116(23):238002, June 2016.
\newblock Publisher: American Physical Society.

\bibitem{charbonneau_glass_2017}
Patrick Charbonneau, Jorge Kurchan, Giorgio Parisi, Pierfrancesco Urbani, and Francesco Zamponi.
\newblock Glass and {Jamming} {Transitions}: {From} {Exact} {Results} to {Finite}-{Dimensional} {Descriptions}.
\newblock {\em Annual Review of Condensed Matter Physics}, 8(1):265--288, 2017.
\newblock \_eprint: https://doi.org/10.1146/annurev-conmatphys-031016-025334.

\bibitem{rintoul_metastability_1996}
M.~D. Rintoul and S.~Torquato.
\newblock Metastability and {Crystallization} in {Hard}-{Sphere} {Systems}.
\newblock {\em Physical Review Letters}, 77(20):4198--4201, November 1996.
\newblock Publisher: American Physical Society.

\bibitem{charbonneau_finite-size_2021}
Patrick Charbonneau, Eric~I. Corwin, R.~Cameron Dennis, Rafael Díaz Hernández~Rojas, Harukuni Ikeda, Giorgio Parisi, and Federico Ricci-Tersenghi.
\newblock Finite-size effects in the microscopic critical properties of jammed configurations: {A} comprehensive study of the effects of different types of disorder.
\newblock {\em Physical Review E}, 104(1):014102, July 2021.

\bibitem{arceri_jamming_2022}
Francesco Arceri, Eric~I. Corwin, and Corey~S. O'Hern.
\newblock The {Jamming} {Transition} and the {Marginally} {Stable} {Solid}, September 2022.
\newblock arXiv:2209.02829 [cond-mat].

\bibitem{bertrand_protocol_2016}
T.~Bertrand, R.~P. Behringer, B.~Chakraborty, C.~S. O'Hern, and M.~D. Shattuck.
\newblock Protocol dependence of the jamming transition.
\newblock {\em Physical Review E}, 93(1):012901, January 2016.
\newblock arXiv:1506.05041 [cond-mat].

\bibitem{artiaco_exploratory_2020}
Claudia Artiaco, Paolo Baldan, and Giorgio Parisi.
\newblock An exploratory study of the glassy landscape near jamming.
\newblock {\em Physical Review E}, 101(5):052605, May 2020.
\newblock arXiv:1908.06127 [cond-mat].

\bibitem{charbonneau_exact_2014}
Patrick Charbonneau, Jorge Kurchan, Giorgio Parisi, Pierfrancesco Urbani, and Francesco Zamponi.
\newblock Exact theory of dense amorphous hard spheres in high dimension. {III}. {The} full {RSB} solution.
\newblock {\em Journal of Statistical Mechanics: Theory and Experiment}, 2014(10):P10009, October 2014.
\newblock arXiv:1310.2549 [cond-mat].

\bibitem{dennis_jamming_2020}
R.C. Dennis and E.I. Corwin.
\newblock Jamming {Energy} {Landscape} is {Hierarchical} and {Ultrametric}.
\newblock {\em Physical Review Letters}, 124(7):078002, February 2020.

\bibitem{charbonneau_fractal_2014-1}
Patrick Charbonneau, Jorge Kurchan, Giorgio Parisi, Pierfrancesco Urbani, and Francesco Zamponi.
\newblock Fractal free energy landscapes in structural glasses.
\newblock {\em Nature Communications}, 5(1):3725, April 2014.
\newblock Publisher: Nature Publishing Group.

\bibitem{thirumalaiswamy_exploring_2022}
Amruthesh Thirumalaiswamy, Robert~A. Riggleman, and John~C. Crocker.
\newblock Exploring canyons in glassy energy landscapes using metadynamics.
\newblock {\em Proceedings of the National Academy of Sciences}, 119(43):e2210535119, October 2022.
\newblock arXiv:2204.00587 [cond-mat].

\bibitem{rainone_following_2015}
Corrado Rainone, Pierfrancesco Urbani, Hajime Yoshino, and Francesco Zamponi.
\newblock Following the {Evolution} of {Hard} {Sphere} {Glasses} in {Infinite} {Dimensions} under {External} {Perturbations}: {Compression} and {Shear} {Strain}.
\newblock {\em Physical Review Letters}, 114(1):015701, January 2015.
\newblock Publisher: American Physical Society.

\bibitem{urbani_shear_2017}
Pierfrancesco Urbani and Francesco Zamponi.
\newblock Shear {Yielding} and {Shear} {Jamming} of {Dense} {Hard} {Sphere} {Glasses}.
\newblock {\em Physical Review Letters}, 118(3):038001, January 2017.
\newblock Publisher: American Physical Society.

\bibitem{sciortino_potential_2005}
Francesco Sciortino.
\newblock Potential energy landscape description of supercooled liquids and glasses.
\newblock {\em Journal of Statistical Mechanics: Theory and Experiment}, 2005(05):P05015, May 2005.

\bibitem{martiniani_when_2023}
Stefano Martiniani and Mathias Casiulis.
\newblock When you can’t count, sample! {Computable} entropies beyond equilibrium from basin volumes.
\newblock {\em Papers in Physics}, 15:150001--150001, February 2023.

\bibitem{speedy_random_1998}
Robin~J. Speedy.
\newblock Random jammed packings of hard discs and spheres.
\newblock {\em Journal of Physics: Condensed Matter}, 10(19):4185, May 1998.

\bibitem{martiniani_structural_2016}
Stefano Martiniani, K.~Julian Schrenk, Jacob~D. Stevenson, David~J. Wales, and Daan Frenkel.
\newblock Structural analysis of high-dimensional basins of attraction.
\newblock {\em Physical Review E}, 94(3):031301, September 2016.
\newblock arXiv:1603.09627 [cond-mat].

\bibitem{gao_enumeration_2007}
Guo-Jie Gao, Jerzy Blawzdziewicz, and Corey~S. O'Hern.
\newblock Enumeration of distinct mechanically stable disk packings in small systems.
\newblock {\em Philosophical Magazine}, 87(3-5):425--431, January 2007.
\newblock arXiv:cond-mat/0605009.

\bibitem{ashwin_calculations_2012}
S.~S. Ashwin, J.~Blawzdziewicz, C.~S. O'Hern, and M.~D. Shattuck.
\newblock Calculations of the {Structure} of {Basin} {Volumes} for {Mechanically} {Stable} {Packings}.
\newblock {\em Physical Review E}, 85(6):061307, June 2012.
\newblock arXiv:1112.4234 [cond-mat].

\bibitem{martiniani_numerical_2017}
Stefano Martiniani, K.~Julian Schrenk, Kabir Ramola, Bulbul Chakraborty, and Daan Frenkel.
\newblock Numerical test of the {Edwards} conjecture shows that all packings are equally probable at jamming.
\newblock {\em Nature Physics}, 13(9):848--851, September 2017.
\newblock Publisher: Nature Publishing Group.

\bibitem{hindmarsh_sundials_2005}
Alan~C. Hindmarsh, Peter~N. Brown, Keith~E. Grant, Steven~L. Lee, Radu Serban, Dan~E. Shumaker, and Carol~S. Woodward.
\newblock {SUNDIALS}: {Suite} of nonlinear and differential/algebraic equation solvers.
\newblock {\em ACM Transactions on Mathematical Software}, 31(3):363--396, September 2005.
\newblock Publisher: Association for Computing Machinery (ACM).

\bibitem{gardner_enabling_2022}
David~J. Gardner, Daniel~R. Reynolds, Carol~S. Woodward, and Cody~J. Balos.
\newblock Enabling {New} {Flexibility} in the {SUNDIALS} {Suite} of {Nonlinear} and {Differential}/{Algebraic} {Equation} {Solvers}.
\newblock {\em ACM Transactions on Mathematical Software}, 48(3):1--24, September 2022.
\newblock arXiv:2011.10073 [cs].

\bibitem{bitzek_structural_2006}
Erik Bitzek, Pekka Koskinen, Franz Gähler, Michael Moseler, and Peter Gumbsch.
\newblock Structural {Relaxation} {Made} {Simple}.
\newblock {\em Physical Review Letters}, 97(17):170201, October 2006.
\newblock Publisher: American Physical Society.

\bibitem{parisi_theory_2020}
Giorgio Parisi, Pierfrancesco Urbani, and Francesco Zamponi.
\newblock {\em Theory of {Simple} {Glasses}: {Exact} {Solutions} in {Infinite} {Dimensions}}.
\newblock Cambridge University Press, 1 edition, January 2020.

\bibitem{liu_limited_1989}
Dong~C. Liu and Jorge Nocedal.
\newblock On the limited memory {BFGS} method for large scale optimization.
\newblock {\em Mathematical Programming}, 45(1):503--528, August 1989.

\bibitem{cauchy_methode_nodate}
M~Augustine Cauchy.
\newblock Me´thode ge´ne´rale pour la re´solution des syste`mes d’e´quations simultane´es.

\bibitem{suryadevara_basins_2025}
Praharsh Suryadevara, Mathias Casiulis, and Stefano Martiniani.
\newblock The {Basins} of {Attraction} of {Soft} {Sphere} {Packings} {Are} {Not} {Fractal}, July 2025.
\newblock arXiv:2409.12113 [cond-mat].

\bibitem{charbonneau_jamming_2023}
Patrick Charbonneau and Peter~K. Morse.
\newblock Jamming, relaxation, and memory in a minimally structured glass former.
\newblock {\em Physical Review E}, 108(5):054102, November 2023.
\newblock Publisher: American Physical Society.

\bibitem{morse_geometric_2014}
Peter~K. Morse and Eric~I. Corwin.
\newblock Geometric {Signatures} of {Jamming} in the {Mechanical} {Vacuum}.
\newblock {\em Physical Review Letters}, 112(11):115701, March 2014.
\newblock Publisher: American Physical Society.

\bibitem{charbonneau_universal_2016}
Patrick Charbonneau, Eric~I. Corwin, Giorgio Parisi, Alexis Poncet, and Francesco Zamponi.
\newblock Universal {Non}-{Debye} {Scaling} in the {Density} of {States} of {Amorphous} {Solids}.
\newblock {\em Physical Review Letters}, 117(4):045503, July 2016.
\newblock Publisher: American Physical Society.

\bibitem{morse_echoes_2017}
Peter~K. Morse and Eric~I. Corwin.
\newblock Echoes of the {Glass} {Transition} in {Athermal} {Soft} {Spheres}.
\newblock {\em Physical Review Letters}, 119(11):118003, September 2017.
\newblock Publisher: American Physical Society.

\bibitem{hagh_order_2024}
Varda~F. Hagh and Sidney~R. Nagel.
\newblock Order in disordered packings with and without permutation symmetry, March 2024.
\newblock arXiv:2403.03926 [cond-mat].

\bibitem{kurchan_phase_1996}
Jorge Kurchan and Laurent Laloux.
\newblock Phase space geometry and slow dynamics.
\newblock {\em Journal of Physics A: Mathematical and General}, 29(9):1929--1948, May 1996.
\newblock arXiv:cond-mat/9510079.

\bibitem{grigera_geometric_2002}
Tomás~S. Grigera, Andrea Cavagna, Irene Giardina, and Giorgio Parisi.
\newblock Geometric {Approach} to the {Dynamic} {Glass} {Transition}.
\newblock {\em Physical Review Letters}, 88(5):055502, January 2002.
\newblock Publisher: American Physical Society.

\bibitem{cavagna_numerical_2004}
Andrea Cavagna, Irene Giardina, and Giorgio Parisi.
\newblock Numerical {Study} of {Metastable} {States} in {Ising} {Spin} {Glasses}.
\newblock {\em Physical Review Letters}, 92(12):120603, March 2004.
\newblock Publisher: American Physical Society.

\bibitem{cavagna_stationary_1998}
Andrea Cavagna, Irene Giardina, and Giorgio Parisi.
\newblock Stationary points of the {Thouless}-{Anderson}-{Palmer} free energy.
\newblock {\em Physical Review B}, 57(18):11251--11257, May 1998.
\newblock Publisher: American Physical Society.

\bibitem{cavagna_fragile_2001}
A.~Cavagna.
\newblock Fragile vs. strong liquids: {A} saddles-ruled scenario.
\newblock {\em Europhysics Letters}, 53(4):490, February 2001.
\newblock Publisher: IOP Publishing.

\bibitem{cavagna_role_2001}
Andrea Cavagna, Irene Giardina, and Giorgio Parisi.
\newblock Role of saddles in mean-field dynamics above the glass transition.
\newblock {\em Journal of Physics A: Mathematical and General}, 34(26):5317, June 2001.

\bibitem{weibull_statistical_1951}
Waloddi Weibull.
\newblock A {Statistical} {Distribution} {Function} of {Wide} {Applicability}.
\newblock {\em Journal of Applied Mechanics}, 1951.
\newblock Publisher: American Society of Mechanical Engineers.

\bibitem{majumdar_extreme_2020}
Satya~N. Majumdar, Arnab Pal, and Grégory Schehr.
\newblock Extreme value statistics of correlated random variables: {A} pedagogical review.
\newblock {\em Physics Reports}, 840:1--32, January 2020.

\end{thebibliography}

\end{document}